# A Possible Role for Entropy in Creative Cognition

**Liane Gabora**
University of British Columbia
liane.gabora@ubc.ca

**Abstract**
This paper states the case for applying the conceptual and analytic tools associated with the study of entropy in physical systems to cognition, focusing on creative cognition. It is proposed that minds modify their contents and adapt to their environments to minimize *psychological entropy*: arousal-provoking uncertainty, which can be experienced negatively as anxiety, or positively as a wellspring for creativity (or both). Thus, intrinsically motivated creativity begins with detection of high psychological entropy material (e.g., a question or inconsistency), which provokes uncertainty and is arousal-inducing. This material is recursively considering from new contexts until it is sufficiently restructured that arousal dissipates and entropy reaches an acceptable level. Restructuring involves neural synchrony and dynamic binding, and may be facilitated by temporarily shifting to a more associative mode of thought. The creative outcome may similarly induce restructuring in others, and thereby contribute to the cultural evolution of more nuanced understandings. Thus, the concept of entropy could play a unifying role in cognitive science as a driver of thought and action, and in cultural studies as the driver of the creative innovations that fuel cultural evolution. The paper concludes with an invitation for cross-disciplinary exploration of this potential new arena of entropy studies.

## 1. Introduction

It has been suggested that minds modify their contents and adapt to their environments to minimize 'psychological entropy' [1]. The concept of entropy, which comes from thermodynamics and information theory, refers to the amount of uncertainty and disorder in a system. Self-organizing systems continually interact with and adapt to their environments to minimize internal entropy. Open systems such as living organisms capture energy (or information) from their environment, use it to maintain semi-stable, far-from-equilibrium states, and displace entropy into the outside world to keep their own entropy low. (The displaced entropy is sometimes called negentropy.)

Hirsh, Mar, and Peterson used the term *psychological entropy* to refer to anxiety-provoking uncertainty, which they claim humans attempt to keep at a manageable level. The concept of psychological entropy was redefined as referring to *arousal*-provoking uncertainty, which can be experienced negatively as anxiety, or positively as a wellspring for creativity (or both) [2]. This redefinition is consistent with findings that creative individuals exhibit greater openness to experience and higher tolerance of ambiguity [3], which could make them more tolerant of uncertainty. It is also consistent with findings that creative individuals exhibit greater variability in arousal level [4]; this could reflect a predisposition to invite situations that increase psychological entropy, experience them positively, and resolve them.

This paper makes the case that the concept of entropy could play a pivotal role in our understanding of cognitive systems as a driver and motivator of thought and action, and in cultural studies as the driver of the creative innovations that fuel cultural evolution. The paper outlines the theoretical basis for applying entropy to the study of cognition in general, and creative cognition in particular. It begins with background material on cognitive science and the psychology of creativity. It then suggests a unifying role for the concept of entropy in this field.

## 2. Background: Cognitive Science and the Study of Creativity

Cognitive science is the interdisciplinary study of the how the mind works. It differs from psychology in that it focuses less on clinical and social aspects, and more on the processes by which information is perceived and processed, and how these processes give rise to intelligence and behavior. The methods used by cognitive scientists include experimental research with human participants and animals, computational models (e.g., neural network models such as deep learning), mathematical models, and neuroscientific approaches such as functional magnetic resonance imaging (fMRI). Research in the cognitive sciences is somewhat fragmented. There are extensive bodies of research on specific cognitive abilities and processes such as attention, perception, mental representation, language, memory, planning, problem solving, emotion, and so forth. However, there is a relative paucity of research into how these abilities work together as a whole.

The study of creativity focuses on cognitive processes that result in outcomes that are new / original and useful / appropriate, or that have a transformative effect on the mind of the creator and/or beholder of a creative work. It is a relatively minor area of cognitive science, but it is an area where the fragmentary nature of cognitive science becomes very apparent, because many if not most of the processes independently studied by cognitive scientists (e.g., attention, mental representation, emotion, et cetera) come together in creative thinking, as well as others more specifically tied to creativity (e.g., intuition, incubation, divergent thinking, and insight). To understand how an engineer redesigns a machine to solve a problem encountered with its predecessor, or how a musician expresses the pain of a breakup in a piece of music, we can no longer treat perception, mental representation, emotion, and so forth as separate. An explanation of how such creative ideas and products come about entails the synthesis of multiple cognitive processes. That is, it must incorporate how the mental representation of a situation or problem guides what the creator pays attention to, and how the creator translates emotion into words and notes. It must tell us why creative self-expression can be intrinsically rewarding [5-7], why it is correlated with positive affect [8], and often therapeutic [9,10], and why it can enhance ones' sense of self [11,12]. It is proposed that the concept of psychological entropy can play an important role in coming to a scientific understanding of how cognitive abilities come together in the execution of complex real-life tasks by serving as a generalized driver of cognition in general, and creative cognition in particular.

**3. A Role for Entropy in Creative Cognition**
  The concept of psychological entropy serves as a starting point for the *honing theory of creativity* (HT) [2,13-15], according to which minds, like biological organisms, are self-organizing, self-maintaining, self-reproducing, entropy-minimizing structures. HT posits that creativity uses psychological entropy—a macro-level variable acting at the level of the mind as a whole—to drive emotions and intuitions that play a role in initiating, tracking, and monitoring creative progress. Just as a wounded organism spontaneously heals, when one encounters a situation that challenges expectations or beliefs, this signals that a particular arena of understanding could benefit from self-organized change.
  Thus, creativity begins with the detection of high psychological entropy material (e.g., a question or inconsistency), which provokes uncertainty and is arousal-inducing. The creative process involves recursively considering this material from new contexts until it is sufficiently restructured that arousal dissipates. Restructuring involves neural synchrony—simultaneous oscillation of membrane potentials in networks of neurons connected by way of electrical synapses—so as to dynamically update the strengths with which different properties, concepts, or ideas are bound together in memory. Restructuring may be facilitated by temporarily defocusing attention and shifting to a more divergent or associative mode of thought, which is conducive to the forging of associations between seemingly disparate but potentially useful or interesting concepts or ideas. Creative restructuring may reduce dissonance, unify previously disparate findings, or facilitate the identification and expression of repressed emotion.
  The creative outcome may similarly induce restructuring in others, and thereby contribute to the cultural evolution of more nuanced understandings. Only those outcomes that induce cognitive restructuring in others and are experienced as useful, aesthetically appealing, or interesting in some way serve as building blocks for others' creative thinking. Because the creative process builds on the outcomes of previous creators' efforts, cultural evolution is cumulative. HT posits that, as did very early life [17-24], culture evolves, not through Darwinian natural selection but through communal exchange amongst self-organizing networks, in this case, minds.
  Thus, psychological entropy not only drives the creative processes of individuals but also the cultural processes that have transformed our planet. Since lines of cultural descent connecting creative outputs can exhibit little continuity (e.g., a piece of music may inspire a painting or book), what is evolving through culture is the self-organizing mind, as opposed to discrete elements of culture (or 'memes') such as songs or sayings [25-27]. Creative outputs reflect the cultural evolutionary state of the worldviews that generate them.

## 3. Discussion

Theories of cognition that do not address how different cognitive processes work together as a whole are ill-equipped to explain the above-mentioned intrinsically rewarding nature of creativity. However, theories that take as a departure point the idea that creative processes—indeed thought itself—is driven by the desire to reduce psychological entropy, are better positioned to explain this aspect of creativity. By restructuring ones' network of concepts, ideas, and attitudes, the creative process reduces the anxiety-provoking uncertainty of psychological entropy, and is therefore experienced as positive and potentially therapeutic.

Interestingly, the transformative effect may operate across domains [28], e.g., one may obtain an understanding of ecological webs by painting murals of them, and likewise, by painting an ecological web get a better understanding of mural painting. Again, an explanation for this requires theoretical constructs such as psychological entropy that transcend specific cognitive processes or domains.

## 4. Conclusions

It is suggested that the domain of application of the concept of entropy can be fruitfully extended from the natural sciences to the cognitive sciences by considering the mind as an intrinsically entropy-minimizing structure, and more specifically, the creativity process as a means by which psychological entropy is minimized. Thus, conceptual and analytic tools associated with the study of entropy in physical systems can be fruitfully applied to cognition. It is hoped that through interdisciplinary collaborations the formalization and development of these nascent ideas will yield a richer understanding of our uniquely innovative capacities.

**Acknowledgments**
This work was supported by a grant (62R06523) from the Natural Sciences and Engineering Research Council of Canada. I have not received funds specifically allocated for covering the costs to publish in open access.